\documentclass[12pt]{article}
\usepackage{amssymb,amsmath,amsthm,latexsym}
\usepackage{authblk}
\usepackage{amsmath,amssymb}
\usepackage[final]{graphics}
\usepackage{color}

\usepackage{enumerate}
\usepackage{latexsym}
\def\nz{\ifmmode {I\hskip -3pt N} \else {\hbox {$I\hskip -3pt N$}}\fi}
\def\zz{\ifmmode {Z\hskip -4.8pt Z} \else
       {\hbox {$Z\hskip -4.8pt Z$}}\fi}
\def\qz{\ifmmode {Q\hskip -5.0pt\vrule height6.0pt depth 0pt
       \hskip 6pt} \'else {\hbox
       {$Q\hskip -5.0pt\vrule height6.0pt depth 0pt\hskip 6pt$}}\fi}
\def\rz{\ifmmode {I\hskip -3pt R} \else {\hbox {$I\hskip -3pt R$}}\fi}

\def\cz{\ifmmode {C\hskip -4.8pt\vrule height5.8pt\hskip 6.3pt} \else
       {\hbox {$C\hskip -4.8pt\vrule height5.8pt\hskip 6.3pt$}}\fi}
\def\tz{\ifmmode {T\hskip -4.8pt\vrule height5.8pt\hskip 6.3pt} \else
       {\hbox {$T\hskip -4.8pt\vrule height5.8pt\hskip 6.3pt$}}\fi}

\def\Ag {{\cal A}} %A gothique
\def\Bg {{\cal B}} %B gothique
\def\Cg {{\cal C}} %C gothique
\def\Cg {{\cal C}} %C gothique
\def\Dg {{\cal D}} %D gothique
\def\Eg {{\cal E}} %E gothique
\def\Fg {{\cal F}}
\def\Gg {{\cal G}} %H gothique
\def\Hg {{\cal H}} %H gothique
\def\Lg {{\cal L}} %L gothique
\def\Mg {{\cal M}}
\def\Ng {{\cal N}} %N gothique
\def\Qg {{\cal Q}} %Q gothique
\def\Sg {{\cal S}} %S gothique
\def\Bb {{\bf B}}
\def\a{\alpha}

\def\c{\gamma}
\def\Ga{\Gamma}
\def\d{\delta}

\def\p{\psi}

\def\ph{\varphi}
\def\ep{\varepsilon}

\def\l{\lambda}
\def\L{\Lambda}

\def\r{\rho}
\def\s{\sigma}
\def\t{\tau}

\def\W{\Omega}

\def\ta{\tilde{a}}

\def\tLg{\tilde{\Lg}}

\def\Ker{{\rm Ker\,}}

\def\lbeq(#1){\label{eqn:#1}}
\def\refeq(#1){{\rm (\ref{eqn:#1})}}
\def\lbsec(#1){\label{s:#1}}
\def\refsec(#1){{\rm \S\ref{s:#1}}}
\def\lbsubsec(#1){\label{ss:#1}}
\def\refsubsec(#1){{\rm \S\ref{ss:#1}}}
\def\lbth(#1){\label{th:#1}}
\def\refth(#1){{\rm Theorem \ref{th:#1}}}
\def\refthb(#1){{\bf Theorem \ref{th:#1}}}
\def\lblm(#1){\label{lm:#1}}
\def\reflm(#1){{\rm Lemma \ref{lm:#1}}}
\def\lbprp(#1){\label{prp:#1}}
\def\refprp(#1){{\rm Proposition \ref{prp:#1}}}
\def\lbass(#1){\label{ass:#1}}
\def\refass(#1){{\rm Assumption \ref{ass:#1}}}
\def\lbcor(#1){\label{cor:#1}}
\def\refcor(#1){{\rm Corollary \ref{cor:#1}}}
\def\lbexm(#1){\label{exm:#1}}
\def\refexm(#1){{\rm Example \ref{exm:#1}}}
\def\lbdf(#1){\label{df:#1}}
\def\refdf(#1){{\rm Definition \ref{df:#1}}}
\def\lbpbs(#1){\label{pbs:#1}}
\def\refpbs(#1){{\rm Problem \ref{pbs:#1}}}

\def\bgdf{\begin{definition}}
\def\eddf{\end{definition}}
\def\bgth{\begin{theorem}}
\def\edth{\end{theorem}}
\def\bglm{\begin{lemma}}
\def\edlm{\end{lemma}}
\def\bgprp{\begin{proposition}}
\def\edprp{\end{proposition}}
\def\bgcor{\begin{corollary}}
\def\edcor{\end{corollary}}
\def\bgexm{\begin{example}}
\def\edexm{\end{example}}
\def\bgpf{\begin{proof}}
\def\edpf{\end{proof}}
\def\bgpbs{\begin{problems}}
\def\edpbs{\end{problems}}
\def\bgrm{\begin{remark}}
\def\edrm{\end{remark}}
\def\bgrms{\begin{remarks}}
\def\edrms{\end{remarks}}
\def\bgass{\begin{assumption}}
\def\edass{\end{assumption}}
\def\ben{\begin{enumerate}}
\def\een{\end{enumerate}}

\def\ep{\varepsilon}

\def\R{{\mathbb R}}
\def\C{{\mathbb C}}
\def\Cb{\overline{{\mathbb C}}}

\def\la{\langle}
\def\bqn{\begin{equation}}
\def\eqn{\end{equation}}
\def\bqan{\begin{eqnarray*}}
\def\eqan{\end{eqnarray*}}
\def\bqa{\begin{eqnarray}}
\def\eqa{\end{eqnarray}}

\def\ra{\rangle}

\def\ax{{\la x \ra}}

\def\ay{{\la y \ra}}
\def\br{\begin{array}}
\def\er{\end{array}}
\def\lap{\Delta}

\newcommand {\absleq} {{\leq_{|\, \cdot\, |}\, }}

\newcommand {\pa}{\partial}


\newtheorem{theorem}{Theorem}[section]
\newtheorem{lemma}[theorem]{Lemma}
\newtheorem{proposition}[theorem]{Proposition}
\newtheorem{definition}[theorem]{Definition}
\newtheorem{remark}[theorem]{Remark}
\newtheorem{corollary}[theorem]{Corollary}
\newtheorem{example}[theorem]{Example}
\newtheorem{assumption}[theorem]{Assumption}
\title{\LARGE \bf On the approximation by regular potentials of 
Schr\"odinger operators with point interactions}
\author{Artbazar Galtbayar and Kenji Yajima}

\date{}

\begin{document}
%\pagenumbering{roman}
\maketitle

\begin{abstract}
We prove that the wave operators for Schr\"odinger operators with 
multi-center local point interactions are the scaling limits of the ones 
for Schr\"odinger operators with regular potentials. We simulataneously 
present a proof of the corresponding well known result for 
the resolvent which substantially simplifies the one by Albeverio et al. 
\end{abstract}
{2010 {\it Mathematics Subject Classification} 47A10, 81Q10, 81Uxx }

\section{Introduction} 
Let $Y=\{y_1,\dots,y_N\}$ be the set of $N$ points in $\R^3$ 
and $T_0$ be the densely defined non-negative symmetric operator 
in $\Hg=L^2(\R^3)$ defined by 
\[
T_0=-\Delta\vert_{C_0^\infty(\R^3 \setminus Y)}.
\]
Any of selfadjoint extensions of $T_0$ is called 
the Schr\"odinger operator 
with point interactions at $Y$. Among them, we are concerned  
with the ones with local point interactions $H_{\a,Y}$ which are defined by 
separated boundary conditions at each point $y_j$ parameterized by 
$\a_j \in \R$, 
$j=1, \dots, N$. 
They can be defined via the resolvent equation (cf. \cite{AHK}): With 
$H_0= -\lap$ being the free Schr\"odinger operator and 
$z \in \C^{+}=\{z\in \C | \Im z >0\}$,  
\bqn \lbeq(resolvent_identity)
(H_{\a,Y} -z^2)^{-1} = (H_0-z^2)^{-1} + 
\sum_{j,\ell=1}^N  (\Ga_{\a,Y}(z)^{-1})_{j\ell} 
\,\Gg_{z}^{y_j} \otimes \overline{\Gg_{z}^{y_\ell}}, 
\eqn 
where $\a=(\a_1, \dots, \a_N) \in \R^N$, 
$\Gamma_{\a,Y}(z)$ is $N\times N$ symmetric 
matrix whose entries are entire holomorphic functions of $z\in \C$ given by 
\bqn \lbeq(ga-def)
\Gamma_{\a,Y}(z)\;:=\;\Big(\Big(\a_j-\frac{iz}{\,4\pi\,}\Big)
\d_{j\ell}-\Gg_z(y_j-y_\ell)\hat{\d}_{j\ell}
\Big)_{\!j,\ell=1,\dots,N}\, ,
\eqn 
where $\delta_{j\ell}=1$ for $j=\ell$ and $\delta_{j\ell}=0$ otherwise; 
$\hat{\delta}_{j\ell}=1-\d_{j\ell}$; 
$\Gg_z(x)$ is the convolution kernel of $(H_0-z^2)^{-1}$: 
\bqn \lbeq(Gs-y)
\Gg_z(x)=\frac{e^{iz|x|}}{\,4\pi |x|\,} \ \ \mbox{and} \ \ 
\Gg_z^{y}(x)=\frac{e^{iz|x-y|}}{\,4\pi |x-y|\,}. 
\eqn 
Since $(H_{\a,Y}-z^2)^{-1} - (H_0 -z^2)^{-1}$ is of rank $N$ 
by virtue of \refeq(resolvent_identity), the wave operators 
$W^{\pm}_{\a,Y}$ defined by the limits 
\bqn \lbeq(limit)
W^{\pm}_{\a,Y}u= \lim_{{ t\to\pm \infty}} 
e^{itH_{\a,Y} }e^{-itH_0} u, \quad  u \in \Hg 
\eqn 
exist and are complete in the sense that 
${\rm Image}\,W^{\pm}_{\a,Y}= \Hg_{ac}$, the absolutely continuous 
(AC for short) subspace of $\Hg$ for $H_{\a,Y}$.  Wave operators 
are of fundamental importance in scattering theory. 

This paper is concerned with the approximation of the 
wave operators $W^{\pm}_{\a,Y}$ by the ones for Schr\"odinger 
operators with regular potentials and generalizes 
a result in \cite{DMSY} for the case $N=1$, which 
immediately implies that $W^{\pm}_{\a,Y}$ are bounded in 
$L^p(\R^3)$ for $1<p<3$, see remarks below \refth(theo-1). 
We also give a proof of the corresponding well known result 
for the resolvent $(H_{\a,Y}-z)^{-1}$ which substantially 
simplifies the one in the seminal monograph \cite{AHK}.

We begin with recalling various properties of $H_{\a,Y}$ 
(see \cite{AHK}):  
\ben
\item[$\bullet$] 
Equation \refeq(resolvent_identity) defines a unique selfadjoint 
operator $H_{\a,Y}$ in the Hilbert space $\Hg= L^2(\R^3)$, 
which is real and local. 
\item[$\bullet$] The spectrum of $H_{\a,Y}$ consists of the 
AC part $[0,\infty)$ and at most $N$ 
non-positive eigenvalues.  Positive eigenvalues are absent. We define 
$\Eg=\{ik\in i\R^{+} \colon -k^2\in \s_p(H_{\a,Y})\}$. 
We simply write $\Hg_{ac}$ and  $P_{ac}$  
respectively for the AC subspace $\Hg_{ac}(H_{\a,Y})$ 
of $\Hg$ for $H_{\a,Y}$ and for the projection $P_{ac}(H_{\a,Y})$ 
onto  $\Hg_{ac}$.  
\item[$\bullet$] $H_{\a,Y}$ may be approximated  by a family 
of Schr\"odinger operators with scaled regular potentials 
\begin{equation} \lbeq(hep)
\overline{H}_Y(\ep)=
-\Delta+\sum_{i=1}^N \frac{\lambda_i(\ep)}{\ep^2}
V_i\left(\frac{x-y_i}{\ep}\right), 
\end{equation}
in the sense that for $z \in \C^{+}$
\bqn \lbeq(epto0)
\lim_{\ep \to 0} (\overline{H}_Y(\ep)-z^2)^{-1}u
=(H_{\a,Y}-z^2)^{-1}u, \quad 
\forall u \in \Hg , 
\eqn 
where $V_j$, $j=1, \dots, N$  are such that 
$H_j= -\lap + V_j(x)$ have threshold resonances at $0$ and 
$\l_1(\ep), \dots, \l_N(\ep)$ are smooth real functions of $\ep$ 
such that $\l_j(0)=1$ and $\l_j'(0)\not=0$ 
(see Theorem 1.1 for more details). 
\een
We prove the following theorem (see Section 4 for the definition of 
the threshold resonance). 

\bgth 
\lbth(theo-1) Let $Y$ be the set of $N$ points 
$Y=\{y_1, \dots, y_N\} $. 
Suppose that: 
\ben 
\item[{\rm (1)}] 
$V_1, \dots, V_N$ are real-valued functions such that for some 
$p<3/2$ and $q>3$,  
\bqn \lbeq(assump-V)
\ax^2 V_j \in (L^p\cap L^q)(\R^3),\quad j=1, \dots, N.
\eqn
\item[{\rm (2)}]  $\l_1(\ep), \dots, \l_N(\ep)$ are real $C^2$ functions 
of $\ep\geq 0$ such that 
\[
\l_j(0)=1, \quad \l_j'(0)\not=0, \quad \forall j=1, \dots, N .
\]
\item[{\rm (3)}] $H_j = -\lap + V_j$, $j=1, \dots, N$ 
admits a threshold resonance at $0$. 
\een
Then, the following statements are satisfied: 
\ben 
\item[{\rm (a)}] 
$\overline{H}_Y(\ep)$ converges in the strong resolvent sense 
as in \refeq(epto0) as $\ep \to 0$ to a Schr\"odinger operator 
$H_{\a, Y}$ with point interactions at $Y$ with certain parameters 
$\a=(\a_1, \dots, \a_N)$ to be specified below . 
\item[{\rm (b)}] Wave operators $W_{Y,\ep}^\pm $ for the pair 
$(\overline{H}_Y(\ep),H_0)$ defined by the strong limits 
\bqn \lbeq(kuroda-ciment)
W_{Y,\ep}^\pm u = 
\lim_{t \to \pm \infty} e^{it\overline{H}_Y(\ep)}e^{-itH_0}u, \quad u \in \Hg 
\eqn 
exist and are complete. $W_{Y,\ep}^\pm$ satisfy  
\bqn \lbeq(wave-converge)
\lim_{\ep \to 0} \| W_{Y,\ep}^{\pm} u - W_{\a,Y}^{\pm}u\|_{\Hg}=0, 
\quad u \in \Hg.
\eqn 
\een
\edth 

Note that H\"older's inequality implies 
$V_j \in L^r(\R^3)$ for all $1\leq r \leq q$  
under the condition \refeq(assump-V). 

\bgrm {\rm 
{\rm (i)} It is known 
that $W_{Y,\ep}^{\pm}$ are  
bounded in $L^p(\R^3)$ for $1<p<3$ (\cite{Y}) and, if $\l_j(\ep)=1$ 
for all $j=1, \dots, N$, $\|W_{Y,\ep}^{\pm}\|_{\Bb(L^p)}$ is independent 
of $\ep>0$ and, 
the proof of \refth(theo-1) shows that \refth(theo-1) holds 
with $\a=0$. It follows by virtue of \refeq(wave-converge) that 
$W_{Y,\ep}$ converges to 
$W_{\a=0,Y}$ weakly in $L^p$ and $W_{\a=0,Y}^\pm$ are bounded in 
$L^p(\R^3)$ for $1<p<3$. Actually, the latter result is known for general 
$\a=(\a_1. \dots, \a_N)$ but its proof is long and complicated 
(\cite{DMSY}). Wave operators satisfy the intertwining property  
\[
f(H_{a,Y})\Hg_{ac}(H_{a,Y})=W_{\a,Y}^{\pm\ast}f(H_0)W_{\a,Y}^{\pm\ast}. 
\]
for Borel functions $f$ on $\R$ and, $L^p$ mapping properties 
of $f(H_{a,Y})P_{ac}(H_{a,Y})$ are reduced to those 
for the Fourier multiplier $f(H_0)$ for a certain range of $p$'s.

\noindent 
{\rm (ii)} If some of $H_j= - \lap + V_j$ have no threshold resonance  
then, \refth(theo-1) remains to hold if corresponding 
points of interactions and parameters $(y_j, \a_j)$ are removed 
from $H_{\a, Y}$. 

\noindent 
{\rm (iii)} 
The first statement is long known (see \cite{AHK}). 
We shall present here a simplified proof, providing in particular 
details of the proof of Lemma 1.2.3 of \cite{AHK} where \cite{H} is 
referred to for ``a tedious but straightforward calculation''
by using a result from \cite{Deift} and a simple matrix formula. 

\noindent
{\rm (iv)} The existence and the completeness of wave operators 
$W_{Y,\ep}^\pm$ are well known (cf. \cite{Ku}).

\noindent 
{\rm (v)}  When $N=1$ and $\a=0$, \refeq(wave-converge) is proved  
in \cite{DMSY}. The theorem is a generalization 
for general $\a$ and $N\geq 2$. 

\noindent
{\rm (vi)} The matrix 
 $\Ga_{\a,Y}(k)$ is non-singular for all 
$k \in (0,\infty)$ by virtue of the selfadjointness of $H_{\a,Y}$ 
and $H_0$. Indeed, if it occurred that $\det\Ga_{\a,Y}(k_0)=0$ for 
some $0<k_0$, then 
the selfadjointness of $H_{\a,Y}$ and $H_0$ implied that 
$\Ga_{\a,Y}(k)^{-1}$ had a simple pole at $k_0$ and 
\begin{multline}
2k_0 {\rm Res}_{z=k_0} (\Ga_{\a,Y}(z)^{-1})_{j\ell}
(\Gg_{z}^{y_j},v)(u,\Gg_{z}^{y_\ell}) \\
= \lim_{z=k_0+i\ep, \ep \downarrow 0}
(z^2-k_0^2)\sum_{j,\ell=1}^N  (\Ga_{\a,Y}(z)^{-1})_{j\ell}
(\Gg_{z}^{y_j},v)(u,\Gg_{z}^{y_\ell})\not=0  \lbeq(contradiction)
\end{multline}
for some $u, v \in C_0^\infty(\R^3)$. However, 
the absence of positive eigenvalues of $H_{\a,Y}$ (see \cite{AHK}, 
pp. 116-117) 
and the Lebesgue dominated convergence theorem imply for all 
$u, v \in C_0^\infty(\R^3)$ that 
\begin{multline*}
\lim_{z=k_0+i\ep, \ep \downarrow 0}
(z^2-k_0^2)((H_{\a,Y}-z^2)^{-1}u, v) \\
= 
\lim_{z=k_0+i\ep, \ep \downarrow 0}
\int_{\R}\frac{2ik_0\ep-\ep^2 }{\mu-(k_0+i\ep)^2}(E(d\mu)u,v)
= (E(\{k_0^2\})u,v)=0 
\end{multline*}
and the likewise for 
$(z^2-k_0^2)((H_0 -z^2)^{-1}u, v)$, 
where $E(d\mu)$ is the spectral projection for $H_{\a, Y}$, 
which contradict \refeq(contradiction). 

}
\edrm

For more about point interactions 
we refer to the monograph \cite{AHK} or the introduction of 
\cite{DMSY} and jump into the proof of \refth(theo-1) 
immediately. We prove \refeq(wave-converge) only for $W^{+}_{Y,\ep}$
as $\overline{H}_{Y}(\ep)$ and $H_{\a, Y}$ are real operators 
and the complex conjugation $\Cg$ changes the direction of the time 
which implies $W^{-}_{Y,\ep} = \Cg W^{+}_{Y,\ep}\Cg^{-1}$. 

We write $\Hg$ for  $L^2(\R^3)$, 
$(u,v)$ for the inner product and $\|u\|$ the norm. 
$u \otimes v$ and $|u\ra \la v|$ indiscriminately denote the 
one dimensional operator 
\[
(u \otimes v) f(x)= |u\ra \la v|f \ra(x) = \int_{\R^3} u(x) 
 \overline{v(y)}f(y) dy.
\]   
Integral operators $T$ and their integral kernels $T(x,y)$ are 
identified. Thus we often say that operator $T(x,y)$ satisfies 
such and such properties and etc. $\Bb_2 (\Hg)$ is the space of Hilbert-Schmidt 
operators in $\Hg$ and 
\[
\|T\|_{HS}= \left(\iint_{\R^3\times \R^3}|T(x,y)|^2 dxdy\right)^{1/2} 
\]
is the norm of $\Bb_2(\Hg)$. $\ax=(1+|x|^2)^{1/2}$ 
and $a \absleq b$ means $|a|\leq |b|$.  
For subsets $D_1$ and $D_2$ of the complex plane $\C$, $D_1 \Subset D_2$ means 
$\overline{D_1}$ is a compact subset of the interior of $D_2$. 
\section{Scaling}
 
For $\ep>0$, we let 
\[
(U_\varepsilon f)(x)=\varepsilon^{-3/2}f(x/\varepsilon). 
\]
This is unitary in $\Hg$ and $H_0 =\ep^2 U_\ep^\ast H_0 U_\ep$. 
We define $H(\ep)$ by 
\bqn \lbeq(scale-1)
H(\ep)=\ep^2 U_\ep^\ast \overline{H}_Y (\ep) U_\ep, \quad 
(\overline{H}_Y (\ep)-z^2)^{-1}
= \ep^{2} U_\ep (H(\ep)-\ep^2 z^2 )^{-1} U_\ep^\ast. 
\eqn
Then, $H(\ep)$ is written as   
\[
H(\ep)=
-\Delta+\sum_{i=1}^N \lambda_i(\ep)
V_i\left(x-\frac{y_i}{\ep}\right) \equiv  -\lap +  V(\ep)
\]
and $W_{Y,\ep}^{\pm}$ are transformed as 
\begin{align}
W_{Y,\ep}^{\pm}& = \lim\limits_{t\to \pm\infty}
U_\ep e^{itH(\ep)/\ep^2} e^{-it H_0/\ep^2} U_\ep^\ast = U_\ep W_Y^{\pm}(\ep)U_\ep^\ast, 
\lbeq(scale-w1) \\
\quad & W_Y^{\pm}(\ep) = \lim\limits_{t\to \pm\infty}
U_\ep e^{itH(\ep)} e^{-it H_0} U_\ep^\ast\, . \lbeq(scale-w2)
\end{align} 
We write the translation operator by $\ep^{-1}y_j$ by  
\[
\tau_{j,\ep}f(x)=f\left(x+\frac{y_j}{\ep}\right), 
\quad j=1,\dots, N.
\]
When $\ep=1$, we simply denote $\tau_{j}= \tau_{j,1}$, $j=1, \dots, N$. 
Then,  
\[
V_j\left(x-\frac{y_j}{\ep}\right)= 
\tau_{j,\ep}^\ast V_j(x)\tau_{j,\ep}.
\]  

\section{Stationary representation} 

The following lemma is obvious and well known: 

\bglm The subspace $\Dg_{\ast}= 
\{u \in L^2 \colon \hat u \in C_0^\infty(\R^3\setminus \{0\})\}$ 
is a dense linear subspace of $L^2(\R^3)$. 
\edlm 

It is obvious that  $\|W_{Y,\ep}^{+}u\|= \|W^{+}_{\a, Y}u\|=\|u\|$ for 
every $u \in \Hg$ and, for proving \refeq(wave-converge) it suffices   
to show that 
\bqn \lbeq(wconv)
\lim_{\ep\to 0} 
(W_{Y,\ep}^{+}u, v)= (W^{+}_{\a, Y}u, v), \quad u, v \in \Dg_\ast.
\eqn 
We express 
$W_{Y,\ep}^{+}$ and $W^{+}_{\a, Y}$ via stationary formulae. 
We recall from \cite{DMSY} the following representation 
formula for $W_{\a,Y}^{+}$.
 
\begin{lemma} Let $u,v \in \Dg_\ast$ and let $\Omega_{j\ell}u$ 
be defined for $j,\ell\in\{1,\dots,N\}$ by 
\bqn \lbeq(Omega_jk)
\frac{1}{\pi i}
\int_{0}^\infty \left(
\int_{\R^3}(\Ga_{\alpha,Y}(-k)^{-1})_{j\ell}\,
\Gg_{-k}(x)\big(\Gg_{k}(y)- \Gg_{-k}(y)\big)u(y) dy \right) k dk. 
\end{equation}
Then,  
\begin{equation} \lbeq(stationary-rep-w)
\langle W^{+}_{\a, Y} u,v\rangle\;=\;\langle u,v\rangle 
+\sum_{j,\ell=1}^N \langle \tau_{j}^{\ast}
\Omega_{j\ell}\tau_{\ell} u,v\rangle .
\end{equation}
Note that for $u \in \Dg_\ast$ the inner integral 
in \refeq(Omega_jk) produces a smooth function of $k\in \R$ 
which vanishes outside the compact set 
$\{|\xi| \colon \xi \in {\rm supp}\, \hat u\}$.   
\end{lemma}

For describing the formula for $W^{+}_{Y,\ep}$ corresponding to  
\refeq(Omega_jk) and \refeq(stationary-rep-w), 
we introduce some notation. 
$\Hg^{(N)}= \Hg \oplus \cdots \oplus \Hg$ is the $N$-fold 
direct sum of $\Hg$. Likewise $T^{(N)}=T\oplus \cdots \oplus T$ for 
an operator $T$ on $\Hg$. For $i=1,\dots, N$ we decompose 
$V_i(x)$ as the product: 
\[
V_i(x)=a_i(x) b_i(x), \quad  
a_i(x)=|V_i(x)|^{1/2}, \ 
b_i(x)=|V_i(x)|^{1/2}{\rm sign}(V_i(x))
\]
where ${\rm sign}\, a=\pm 1 $ if $\pm a>0$ and 
${\rm sign}\, a=0$ if $a=0$. We use matrix notation for 
operators on $\Hg^{(N)}$. Thus, we define  
\[
A= 
\begin{pmatrix}
a_{1} &  \cdots      & 0  \\ 
\vdots         & \ddots & \vdots  \\ 
0    &  \cdots  & a_{N} 
\end{pmatrix}, \ \  
B= 
\begin{pmatrix}
b_{1} &  \cdots      & 0  \\ 
\vdots         & \ddots & \vdots  \\ 
0    &  \cdots  & b_{N} 
\end{pmatrix}, 
\ \  
\Lambda(\ep) = 
\begin{pmatrix}
\lambda_1(\ep) &  \cdots      & 0  \\ 
\vdots         & \ddots & \vdots  \\ 
0    &  \cdots  & \lambda_N(\ep) \\
\end{pmatrix}. 
\]
Since $a_j, b_j$ and $\l_j(\ep)$, $j=1, \dots, N$ are real valued,  
multiplications with $A, B$ and $\Lambda(\ep)$ are selfadjoint operators 
on $\Hg^{(N)}$. We also define  the operator $\t_\ep$  by 
\[
 \t_\ep  \colon \Hg \ni f \mapsto \t_\ep f = \begin{pmatrix}
\t_{1,\ep}  f\\ 
\vdots \\  
\tau_{N,\ep}f \\
\end{pmatrix}  \in \Hg^{(N)}
\]
so that  
\[
V(\ep)= 
\sum_{j=1} ^N \lambda_j(\ep) V_j\left(x-\frac{y_j}{\ep}\right) 
= \t_\ep^\ast A  \Lambda(\ep)B \tau_\ep .
\]
We write for the case $\ep=1$ simply as $\tau= \tau_1$ as previously. 
For $z \in \C$, $G_0(z)$ is the integral operator defined by 
\[
G_0(z)u(y) = \frac1{4\pi}\int_{\R^3}\frac{e^{iz |x-y|}}{|x-y|}u(y) dy. 
\]
It is a holomorphic function of $z\in \C^{+}$ with values in $\Bb(\Hg)$ and 
\[
G_0(z)= (H_0-z^2)^{-1}, \ \ \mbox{for} \ \ z \in \C^{+} 
\]
and, it can be extended to various subsets of $\C^{+}$ when considered as 
as a function with values in a space of operators between suitable function spaces.
We also write 
\[
G_\ep(z)= (H(\ep) -z^2)^{-1}
 \ \mbox{for}\ z\in \C^{+} \setminus \{z \colon z^2 \in \s(H(\ep))\}
\]

\bglm \lblm(ab-L) Let $V_1, \dots, V_N$ satisfy the assumption 
\refeq(assump-V) and $z \in \Cb^{+}$. Then: 
\ben 
\item[{\rm (1)}] $a_i, b_j \in L^{2}(\R^3),
\ \ i,j=1, \dots, N$. 
\item[{\rm (2)}]  $a_i G_0 (z) b_j \in \Bb_2(\Hg)$, $1\leq i, j \leq N$. 
\een
\edlm 
\bgpf (1) We have $a_i, b_j \in L^{2}(\R^3)$ for 
$V_j \in L^1(\R^3)$ as was remarked below \refth(theo-1). 

(2) We also have $|a_j|^2=|b_j|^2= |V_j| \in L^{3/2}(\R^3)$ and 
$|x|^{-2}\in L^{3/2, \infty}(\R^3)$. It follows by the generalized Young inequality that 
\[
\iint_{\R^3\times \R^3}\frac{|a_i(x)|^2|b_j(y)|^2}{|x-y|^2} dx dy 
\leq C \|V_i\|_{L^{3/2}}\|V_j\|_{L^{3/2}}. 
\]
Hence, $a_i G_0 (z) b_j$ is of Hilbert-Schmidt type in $L^2(\R^3)$. 
\end{proof} 

Using this notation, we have form 
\refeq(stationary-rep-w) that 
\bqn 
(W^{+}_{\a,Y}u,v)=(u,v) + 
\left\la \big(\Omega_{j\ell}\big)\tau^\ast {u},  \tau^\ast {v}\right\ra_{\Hg^{(N)}}. 
\eqn 

The resolvent equation for $H(\ep)$ may be written as 
\[
G_\ep(z)-G_0(z)= 
- G_0(z)\tau_{\ep}^\ast A  \L(\ep) B \tau_\ep G_\ep (z) 
\]
and the standard argument (see e.g. \cite{RS}) yields    
\bqn 
G_\ep(z) =G_0(z) - 
 G_0(z)\tau_\ep^{\ast}A 
(1+ \L(\ep) B \tau_\ep G_0(z)\tau_\ep^\ast A )^{-1}
\L(\ep) B \t_\ep G_0(z).
\lbeq(resol-eq)
\eqn 
Note that $\tau_\ep R_0(z)\tau_\ep^\ast \not= R_0(z)$ in general 
unless $N=1$. 

Under the assumption \refeq(assump-V) on $V_1, \dots, V_N$ 
the first two statements of the following lemma follow from the 
limiting absorption principle for the free Schr\"odinger operator 
(\cite{Agmon}, \cite{Kuroda}) and the last from the absence of positive 
eigenvalues for $H(\ep)$ (\cite{Koch-Tataru}). 
In what follows we often write $k$ for $z$ when we want emphasize that 
$k$ can also be real. 

\begin{lemma} \lblm(M-de)
 Suppose that $V_1, \dots V_N $ satisfy the 
assumption of \refth(theo-1). Let $0<\ep\leq 1$. Then:
\ben 
\item[{\rm (1)}] For $u \in \Dg_\ast$, 
$\lim_{\d\downarrow 0}
\sup_{k\in \R}\|A \t_\ep G_0(k+i\d) u - A \t_\ep G_0(k)  u\|_{\Hg^{(N)}}=0 $. 
\item[{\rm (2)}] $\lim_{\d\downarrow 0} 
\sup_{k\in \R} 
\|\L(\ep) A \tau_\ep (G_0(k+i\d)-G_0(k)) 
\tau_\ep^\ast A\|_{\Bb(\Hg^{(N)})}=0 $.
\item[{\rm (3)}] Define for $k\in \Cb^{+}=\{k\in \Im k\geq 0\}$, 
\bqn \lbeq(M-def)
M_\ep (k)=\L(\ep) B \tau_\ep G_0(k)\tau_\ep^\ast A.
\eqn 
Then, $M_\ep(k)$ is a compact operator on $\Hg^{(N)}$ and 
$1+M_\ep (k)$ is invertible for all $k\not=0$. $(1+M_\ep(k))^{-1}$ is 
a locally H\"older continuous function of $\Cb^{+}\setminus\{0\}$ 
with values in $\Bb(\Hg^{(N)})$.  
\een
Statements {\rm (1)} and {\rm(2)} remain to hold when 
$A$ is replaced by $B$. 
\end{lemma}

The well known stationary formula for wave operators 
(\cite{Kuroda}) and the resolvent equation \refeq(resol-eq) yield 
\begin{align} \lbeq(s-1) 
& (W_Y ^{+}(\ep) u,v)- (u,v)\\
&= -\frac1{\pi{i}}
\int_{0}^\infty  
\left( (1+ M_\ep(-k))^{-1}\L(\ep) B \t_\ep 
\{G_0(k)-G_0(-k)\}u,  A\t_\ep G_0(k)v\right) k dk. \notag
\end{align} 
For obtaining the corresponding formula for $W_{Y,\ep}^{+}$, 
we scale back \refeq(s-1)  
by using the identity   \refeq(scale-w1) and \refeq(scale-w2). 
Then 
\[
\tau_\ep  U_\ep^\ast  = U_\ep^\ast \tau, 
\]
and change of variable $k$ to $\ep{k}$ produce the 
first statement of the following lemma. Recall $\tau=\tau_{\ep=1}$. 
The second formula is proven in parallel with the first 
by using \refeq(scale-1). 
The following lemma should need no proof. 

\begin{lemma} \lblm(formula-wr)
{\rm (1)} For $u, v \in \Dg^\ast$, we have 
\begin{align}
& (W^{+}_{Y,\ep}u,v) = (u,v) - 
\frac{\ep^{2}}{\pi {i}}\int_{0}^\infty k dk  \lbeq(fr-w)\\
& \left((1+ M_\ep(-{\ep}k))^{-1}\L(\ep)B
\{G_0(k\ep)-G_0(-k\ep)\}^{(N)}U_{\ep}^\ast \tau{u}, 
 A G_0(k\ep)^{(N)}U_\ep^\ast \tau  {v}\right). \notag
\end{align}
{\rm(2)} For $k\in \C^{+}$ with sufficiently large $\Im k$, 
\begin{align} 
& (\overline{H}_Y(\ep)-k^2)^{-1} = G_0(k)  \notag \\
& \quad -\ep^2\tau^\ast U_\ep G_0(k\ep)^{(N)}A(1+M_\ep (\ep{k}))^{-1}
\L(\ep)BG_0(k\ep)^{(N)}U_\ep^\ast \tau ,  \lbeq(fr-w1) 
\end{align}
where $G_0(\pm k\ep)^{(N)}= G_0(\pm k\ep) \oplus \cdots \oplus G_0(\pm k\ep)$ is 
the $N$-fold direct sum of $G_0(\pm k\ep)$. 
\end{lemma}

Notice that for 
$u \in \Dg_\ast$, 
$\{G_0(k\ep)-G_0(-k\ep)\}^{(N)} U_{\ep}^\ast \tau{u}\not=0$  
for $R^{-1}<k <R$ for some $R>0$  and the integral on 
the right of \refeq(fr-w) is only over $[R^{-1},R] \subset (0,\infty)$ 
uniformly for $0<\ep<1$.  
Indeed, if $u\in \Dg_\ast$ and $\hat{u}(\xi)=0$ unless 
$R^{-1} \leq |\xi| \leq R$ for some $R>1$, then, since the translation 
$\tau$ 
does not change the support of $\hat u(\xi/\ep)$, we have  
\[
\Fg (U_{\ep}^\ast \tau {u})(\xi) = 
\ep^{-\frac32}\Fg(\tau {u})\left(\frac{\xi}{\ep}\right)
=0 
\]
unless $R^{-1}\ep\leq  |\xi|\leq R\ep$ and 
\[
\{G_0(k\ep)-G_0(-k\ep)\}U_{\ep}^\ast \tau {u}= 
2i\pi \d(\xi^2-k^2\ep^2)\Fg (U_{\ep}^\ast \tau {u})(\xi)=0 .
\]
for $k >R$ or $k<R^{-1}$.

\section{Limits as $\ep\to 0$}

We study the small $\ep>0$ behavior of the right hand 
sides of \refeq(fr-w) and \refeq(fr-w1). 
For \refeq(fr-w), the argument above shows that we need only 
consider the integral over a compact set 
$K\equiv [R^{-1},R]\subset \R$ {\it which will 
be fixed in this section}. 
Splitting $\ep^2 = \ep \cdot \ep^{1/2} \cdot \ep^{1/2}$ in front of 
the second term on the right, we place 
one $\ep^{1/2}$ each in front of 
$BG_0(\pm k\ep)^{(N)}U_\ep^\ast $ and  
$A G_0(\pm k\ep)^{(N)}U^\ast $ or $U_\ep G_0(k\ep)^{(N)}A$ 
and the remaining $\ep$ in front of $(1 + M_\ep(\pm{\ep}k))^{-1}$. We begin with the following lemma. 
Recall the definition \refeq(Gs-y) of $\Gg_k$.

\bglm \lblm(endterms)
Suppose $a \in L^2(\R^3)$. Then, following  statements are satisfied: 
\ben 
\item[{\rm (1)}] Let $u \in \Dg_\ast $. Then, uniformly in $k \in K$, we have 
\bqn  \lbeq(7-1)
\lim_{\ep\to 0} \|\ep^{\frac12} a G_0(\pm k\ep)U_\ep^\ast u  
- |a \ra \la \Gg_{\pm k}, u \ra \|_{L^2} =0 . 
\eqn 

\item[{\rm (2)}] Let $u \in L^2(\R^3)$. Then, uniformly on compacts of $k \in\C^{+}$, 
we have 
\bqn \lbeq(7-2)
\|\ep^{\frac12} a G_0(k\ep)U_\ep^\ast u\|_{L^2} 
\leq  C (\Im k)^{-1/2} \|a\|_{L^2} \|u\|_{L^2} 
\eqn 
and the convergence \refeq(7-1) with $k$ in place of $\pm k$.   

\item[{\rm (3)}] 
Let $u \in L^2(\R^3)$. Then, uniformly on compacts of $k \in\C^{+}$, 
we have 
\bqn  \lbeq(7-1a)
\lim_{\ep\to 0} \|\ep^{\frac12} U_\ep G_0(k\ep) a u  
- |\Gg_{k} \ra \la a, u \ra \|_{L^2} =0 . 
\eqn 
\een
\edlm 
\bgpf (1) We prove the $+$ case only. The proof for the $-$ case is similar. 
We have $u \in \Sg(\R^3)$ and   
\[
\ep^{\frac12} G_0(k\ep)U_\ep^\ast u(x) 
= 
\frac1{4\pi} \ep^{2} \int_{\R^3} 
\frac{e^{ik\ep |x-y|}}{|x-y|}u(\ep y) dy 
= \frac1{4\pi}\int_{\R^3} \frac{e^{ik|y|}}{|y|}u(y+\ep x) dy.
\]%eqn 
It is then obvious for any $R>0$ and a compact $K \subset \R$ that 
\bqn \lbeq(6-1)
\lim_{\ep \to 0} \sup_{|x|\leq R, k\in K} 
|\ep^{\frac12} G_0(k\ep)U_\ep^\ast u(x) - \la \Gg_k, u \ra |= 0 
\eqn 
Moreover, H\"older's inequality in Lorentz spaces implies that 
\bqn \lbeq(un)
|\la \Gg_k, u \ra|+ \|\ep^{\frac12} G_0(k\ep)U_\ep^\ast u\|_{\infty} 
\leq \|(4\pi |x|)^{-1}\|_{3,\infty}\|u\|_{\frac32,1}.
\eqn 
It follows from \refeq(6-1) that for any $R>0$ 
\bqn \lbeq(Rgeq)
\lim_{\ep \to 0} \sup_{k \in K} \|\ep^{\frac12} a G_0(k\ep)U_\ep^\ast u  
- a \la \Gg_k, u \ra \|_{L^2(|x|\leq R)}=0 
\eqn 
and, from \refeq(un) that 
\begin{align} 
& \|\ep^{\frac12} a G_0(k\ep)U_\ep^\ast u  
- a \la \Gg_k, u \ra \|_{L^2(|x|\geq R)} \notag \\ 
& \qquad \leq 
2\|a\|_{L^2(|x|\geq R)}\|(4\pi |x|)^{-1}\|_{3,\infty}\|u\|_{\frac32,1}
\to 0 . \lbeq(6-2) 
\end{align} 
Combining \refeq(6-1) and \refeq(6-2), we obtain \refeq(7-1) 
for $u \in \Dg_\ast$. (Since $\Dg(\R^3)$ is dense in $L^{3,1}(\R^3)$, 
\refeq(7-1) actually holds for $u \in L^{\frac32,1}(\R^3)$.)   \\[5pt]
(2)  We have 
\[
\|aG_0(k\ep)\|_{HS}^2 = 
\int_{\R^3\times \R^3} \frac{|a(x)|^2 e^{-2{\Im k}\ep|x-y|}}{16|x-y|^2}dx dy 
\leq C (\Im k \ep)^{-1}\|a\|_{L^2}^2.
\]
This implies \refeq(7-2) as $U_\ep^\ast$ is unitary in $L^2(\R^3)$ 
and it suffices to prove the strong convergence on $L^2$ for 
$u \in C_0^\infty(\R^3)$. This, however, follows as in the case (1).\\[5pt] 
(3) We have  
\[
\ep^{\frac12} (U_\ep G_0(k\ep) a u)(x)
=\int_{\R^3} \frac{e^{ik|x-\ep{y}|}}{4\pi|x-\ep{y}|}a(y) u(y)dy 
\]
and Minkowski's inequality implies 
\bqn \lbeq(mink)
\|\ep^{\frac12} U_\ep G_0(k\ep) a u - 
|\Gg_{k} \ra \la a, u \ra\|\leq 
\int_{\R^3}\|\Gg_k(\cdot-\ep{y})-\Gg_k\|_{L^2(\R^3)}|a(y)u(y)|dy .
\eqn  
Plancherel's and Lebesgue's dominated convergence theorems imply 
that for a compact subset $\tilde K$ of $\C^{+}$
\begin{align*}
& \sup_{k\in \tilde K} \|\Gg_k(\cdot+\ep{y})-\Gg_k\| 
= \sup_{k\in \tilde K} \|(\Fg^{-1}\Gg_k)(\xi)(e^{\ep{y}\xi}-1)\|_{L^2(\R^3_\xi)} \\
& =  
\left( 
\int_{\R^3} \sup_{k\in \tilde K}|(|\xi|^2- k^2)^{-1}(e^{i\ep{y}\xi}-1)|^2 
d\xi\right)^{\frac12}
\leq C \left( 
\int_{\R^3} \langle \xi\rangle^{-4}|(e^{i\ep{y}\xi}-1)|^2 d\xi\right)^{\frac12}
\end{align*}
is uniformly bounded for $y\in \R^3$ and converges to $0$ as $\ep \to 0$. 
Thus,  \refeq(7-1a) follows from \refeq(mink) by applying 
Lebesgue's dominated convergence theorem .  
\edpf 

We next study $\ep (1 + M_\ep({\ep}k))^{-1}$ for $\ep \to 0$ 
and $k\in \Cb^{+}\setminus\{0\}$. 
We decompose 
$M_\ep(k)= \L(\ep)B \tau_\ep G_0(\ep k)\tau^\ast_{\ep}A$ 
into the diagonal and the off-diagonal parts: 
\bqn \lbeq(decomposition) 
M_\ep(k) = D_\ep(\ep{k}) + \ep E_\ep(\ep{k})
\eqn  
where the diagonal part is given by 
\bqn \lbeq(diag)
D_\ep(\ep{k})= \begin{pmatrix}
\l_1(\ep)b_1 G_{0}(\ep{k})a_1 &  \cdots      & 0  \\ 
\vdots               & \ddots        & \vdots  \\ 
0                    &  \cdots       & \l_N(\ep) b_N G_0(\ep{k})a_N  
\end{pmatrix} 
\eqn 
and, the off diagonal part 
$\ep E_\ep(\ep{k})= \left(\l_i(\ep)b_i \t_{i,\ep}G_0(\ep{k})
\t_{j,\ep}^\ast a_j\hat\delta_{ij}\right)$ by  
\bqn \lbeq(off-diag)
\ep E_\ep(\ep{k})= 
\ep \left(\lambda_{i}(\ep)\frac{b_i(x)e^{ik|\ep(x-y)+y_i-y_j|}a_j(y)}
{4\pi |\ep(x-y)+y_i-y_j|}\hat{\delta}_{ij}\right)_{ij}.
\eqn

We study $E_\ep(\ep{k})$ first. 
Define constant matrix $\hat{\Gg}(k)$ by     
\[
\hat{\Gg}_{ij}(k)= \Gg_{ij}(k)\hat{\d}_{ij}, \quad 
\Gg_{ij}(k)= \frac1{4\pi}\frac{e^{ik|y_i-y_j|}}{|y_i-y_j|}, 
\quad i\not=j. 
\]
\bglm \lblm(E-estimate) Assume \refeq(assump-V) and let $\W \subset \Cb^{+}$ 
be compact. We have uniformly for $k \in \W$ that   
\bqn \lbeq(off-conv)
\lim_{\ep\to 0} \|E_{\ep}(\pm \ep{k})- 
|B\ra \hat\Gg(\pm k)\la A|\|_{\Bb(\Hg^{(N)})} = 0.
\eqn
$|B\ra \hat\Gg(\pm k)\la A|$ is an operator of rank at most 
$N$ on $\Hg^{(N)}$:  
\[
|B\ra \hat\Gg(\pm k)\la A| 
\equiv \left( b_i(x)\Gg_{ij}(\pm k)a_j(y) \hat{\d}_{ij}\right).
\]
\edlm 
\bgpf We prove the $+$ case only. The $-$ case may be proved similarly. 
Let $k\in K$. Then, 
\begin{align} 
& \left| \frac{e^{ik|\ep(x-y)+y_i-y_j|}}{|\ep(x-y)+y_i-y_j|} 
-\frac{e^{ik|y_i-y_j|}}{|y_i-y_j|} \right| \notag \\
& \leq \frac{|k||\ep(x-y)|}{|\ep(x-y)+y_i-y_j|}+
\frac{|\ep(x-y)|} {|\ep(x-y)+y_i-y_j||y_i-y_j|} \lbeq(off-dai-a)\\
& 
\leq \frac{C |x-y|}{|(x-y)+(y_i-y_j)/\ep|} \lbeq(off-dai)
\end{align} 
for a constant $C>0$ and  we may estimate as 
\begin{align*}
& \|(E_{\ep,ij}(\ep{k})- \l_i(\ep)b_i \Gg_{ij}(k)a_j) u\|_{L^2}
\leq C \left\|\int_{\R^3}  
\frac{|{b_i(x)}|x-y|{a_j(y)}u(y)|}
{|(x-y)+(y_i-y_j)/\ep|} dy \right\| \notag  \\
& \leq 
C \left\|\int_{\R^3}  
\frac{|\ax{b_i(x)}\ay{a_j(y)}u(y)|}
{|(x-y)+(y_i-y_j)/\ep|} dy \right\| \notag \\
& = 
C \left\|\int_{\R^3}  
\frac{|\t_{i,\ep}(\ax{b_i})(x)\t_{j,\ep}(\ay{a_j}u)(y)|}
{|x-y|} dy \right\|. 
\end{align*}
Since the convolution with the Newton potential $|x|^{-1}$ maps 
$L^\frac{6}{5}(\R^3)$ to $L^{6}(\R^3)$ by virtue of 
Hardy-Littlewood-Sobolev's inequality, H\"older's inequality implies 
that the right hand side is bounded by 
\begin{multline}
C\|\ax{b_i}\|_{L^3} \|\ay{a_j} u\|_{L^{6/5}} \\
\leq 
C \|\ax{b_i}\|_{L^3} \|\ax {a_j}\|_{L^{3}} \|u\|_{L^2} 
= C \|\ax^2 V_i\|_{L^{\frac32}}^{\frac12}
\|\ax^2 V_j\|_{L^{\frac32}}^{\frac12}
\|u\|_{L^2}.
\lbeq(HLS)
\end{multline} 
Let $B_R(0)= \{x\colon |x|\leq R\}$ for 
an $R>0$. Then, for 
$\ep>0$ such that $4R \ep< \min |y_i-y_j|$, we have  
\[
\refeq(off-dai-a) \leq 4C\ep, \quad \forall x, y \in B_R(0).
\]
Thus, if $V_j\in C_0^\infty(\R^3)$, $j=1, \dots, N$ are supported by  
$B_R(0)$, then
\[
\|E_{\ep}(\ep{k})- \L(\ep)B \hat{\Gg}(k)A\|_{\Bb(\Hg^{(N)})}\leq 
4C\ep \sum_{j=1}^N \|V_j \|_{L^1}\xrightarrow{\ep \to 0} 0.
\]
Since $C_0^\infty(\R^3)$ 
is a dense subspace of the Banach space 
$(\ax^{-2}L^{3/2}(\R^3)) \cap L^1(\R^3)$, \refeq(HLS) implies 
$\|E_{\ep}(\ep{k})- \L(\ep)B\hat{\Gg}(k)A\|_{\Bb(\Hg^{(N)})}\to 0$ 
as $\ep \to 0$ for general $V_j$'s which satisfies the assumption 
\refeq(assump-V). 
The lemma follows because 
$\L(\ep ) $ converges to the identify matrix. 
\edpf

We have shown in \reflm(ab-L) that  
$b_i G_0(k\ep) a_j $ is of Hilbert-Schmidt type for $k \in \Cb^{+}$ 
and it is well known that $1+ \l_j(\ep)b_j G_0(k\ep) a_j $
is an isomorphism of $\Hg$ unless 
$k^2\ep^2$ is an eigenvalue of $H_j(\ep)=-\lap + \l_j(\ep)V_j$ 
(see \cite{IS}). 
Hence, the absence of 
positive eigenvalues for $H_j(\ep)$ (see e.g. \cite{Koch-Tataru}) implies that 
 $1+ \l_j(\ep)b_j G_0(k\ep) a_j $ is an isomorphism in $\Hg$ for all 
$k \in \Cb^{+} \setminus (\ep^{-1}i\Eg_j(\ep) \cup \{0\})$ where 
$\Eg_j(\ep)=\{k>0 \colon - k^2 \in \s_p(H_j(\ep))\}$. 
Thus, if we fix a compact set $\W \subset \Cb^{+} \setminus \{0\}$.  
$1+D_\ep(\ep{k})$ is invertible in $\Bb(\Hg^{(N)})$ for small $\ep>0$ 
and ${k}\in \W$ and 
\[
1+ M_\ep({\ep}k)=(1+D_\ep(\ep{k}))(1+\ep(1+D_\ep(\ep{k}))^{-1}
E_\ep(\ep{k})).
\]
It follows that  
\bqn \lbeq(MCinv)
(1+ M_\ep({\ep}k))^{-1}=(1+\ep(1+D_\ep(\ep{k}))^{-1}
E_\ep(\ep{k}))^{-1}(1+D_\ep(\ep{k}))^{-1}. 
\eqn 
and we need study the right hand side of \refeq(MCinv) as $\ep \to 0$.   

We begin by studying $\ep(1+D_\ep(\ep{k}))^{-1}$ and,  
since $1+D_\ep(\ep{k})$ is diagonal, we may do it component-wise. 
We first study the case $N=1$. 

\subsection{Threshold analysis for the case $N=1$} 

When $N=1$, we have $M_{\ep}(\ep{k})=D_\ep(\ep{k})$. 

\bglm \lblm(expansion) Let $N=1$, $a=a_1$ and etc and, let 
$\W$ be compact in $\Cb^{+} \setminus \{0\}$. 
Then, for any $0<\r<\r_0$, $\r_0=(3-p)/2p>1/2$, we have following expansions 
in $\W$ in the space of Hilbert-Schmidt operators ${\Bb}_2(\Hg)$:
\begin{gather}
bG_0(k\ep)a=bD_0a + ik\ep bD_1a+ O((k\ep)^{1+\r}), \lbeq(expansion-0) \\  
M_{\ep}(\ep{k})= b D_0 a + \ep\big(\l'(0) b D_0 a + ik b D_1 a \big) 
+ O(\ep^{1+\r})  \lbeq(luG) \\
D_0 =\frac{1}{4\pi|x-y|}, \ \  D_1 =\frac1{4\pi}. 
\end{gather}
where $O((k\ep)^{1+\r})$ and $O(\ep^{1+\r})$ are 
$\Bb_2(\Hg)$-valued functions of $(k,\ep)$ such that  
\[
\|O((k\ep)^{1+\r})\|_{HS}\leq C |k\ep|^{1+\r}, \quad 
\|O(\ep^{1+\r})\|_{HS}\leq C |\ep|^{1+\r}, \quad 0<\ep <1, \ k \in \W.
\]
\edlm 
\bgpf  Since $\Im k \geq 0$ for $k \in \W$, Taylor's formula and 
the interpolation imply that for any $0\leq\r \leq 1$ there exists a 
constant $C_\r>0$ such that 
\[
|e^{ik\ep|x-y|}-\left( 1 + ik \ep |x-y|\right)|\leq C_\r |\ep{k}|^{1+\r}
|x-y|^{1+\r}.
\]
Hence 
\[
\left|D_\ep(\ep{k})(x,y)- \frac{b(x)a(y)}{4\pi |x-y|}-
ik\ep\frac{b(x)a(y)}{4\pi}\right|
\leq C_\r |k|^{1+\r}\ep^{1+\r}|x-y|^\r |b(x)a(y)|.
\]
We have shown in \reflm(ab-L) that 
$D_\ep(\ep{k})$ and $bD_0a$ are Hilbert-Schmidt operators 
and $b D_1 a$ is evidently so as $a, b \in L^2(\R^3)$ (see the remark below 
\refth(theo-1)). As $\ax b(x) , \ay a(y) \in L^{2p}(\R^3)$, we have 
$\ax^{\r}a(x), \ax^{\r}a(y)\in L^2(\R^3)$ for $\r<\r_0$, 
and 
\[
\iint_{\R^3\times\R^3}|x-y|^{2\r} |b(x)a(y)|^2dxdy 
\leq C \|\ax^{\r}b(x)\|_{L^2}^2 \|\ay^{\r}a(y)\|_{L^2}^2 
\]
This prove estimate \refeq(expansion-0). \refeq(luG) follows from 
\refeq(expansion-0) and Taylor's expansion of $\l(\ep)$. 
This completes the proof of the lemma. 
\edpf 

We define 
\bqn \lbeq(defQQ)
Q_0  = 1 + b D_0 a, \quad Q_1 =\l'(0) b D_0 a + ik b D_1 a, 
\quad bD_1 a= (4\pi)^{-1}|b\ra \la a|.
\eqn 

\paragraph{The regular case} 
\bgdf $H=-\lap + V(x)$ is said to be of regular type at $0$ if $Q_0$ 
is invertible in $\Hg$. 
It is of exceptional type if otherwise.  
\eddf 

\bglm \lblm(regular-type) Suppose $N=1$ and that 
$H= -\lap + V(x)$ is of regular type at $0$. Let $\W$ be a compact subset 
of $\Cb^{+}$. Then,  
\bqn \lbeq(3-1)
\lim_{\ep \to 0}\sup_{k \in \W} 
\|\ep(1+ M_\ep({\ep}k))^{-1}\|_{\Bb(\Hg)}=0 .
\eqn 
\edlm 
\bgpf Since $Q_0 = 1+ bD_0 a$ is invertible, 
\refeq(luG) implies the same for 
$1+M_\ep(\ep{k})$ for $k \in \W$ and small $\ep>0$ and,  
\[
\lim_{\ep \to 0}\sup_{k \in \W}
\|(1+M_\ep(\ep{k}))^{-1}- Q_0^{-1}\|_{\Bb(\Hg)}=0.
\]
\refeq(3-1) follows evidently.  
\edpf
An application of \reflm(M-de), \reflm(endterms) 
and \reflm(regular-type) to \refeq(fr-w) and \refeq(fr-w1) 
immediately produces following proposition for the case $N=1$. 
\bgprp Suppose $H= -\lap + V$ is of regular type at $0$. Then: 
\ben 
\item[{\rm (1)}] As $\ep\to 0$, $ W^{+}_{Y,\ep} =1$ converges strongly 
to the identity operator.  
\item[{\rm (2)}] Let $\W_0\subset \Cb^{+}$ be compact. Then, 
$a(\overline{H}_Y(\ep)-k^2)^{-1}b - aG_0(k)b \to 0$ 
in the norm of $\Bb(\Hg)$  
as $\ep \to 0$ uniformly with respect to $k\in \W_0$

\item[{\rm (3)}] Let $\W_1 \Subset \C^{+}$. Then, 
$\lim\limits_{\ep \to 0}
\sup\limits_{k\in \W_1}\|(\overline{H}_Y(\ep)-k^2)^{-1} - G_0(k)\|_{\Bb(\Hg)}=0$.
\een 
\edprp 

\paragraph{Exceptional case}  Suppose next that $Q_0$ is {\it not} invertible 
and define 
\[
\Mg=\colon {\rm Ker}\, Q_0, \quad  
\Ng=\Ker Q_0^\ast, \quad Q_0^\ast = 1+ aD_0 b.
\]
By virtue of the Riesz-Schauader theorem   
$\dim \Mg= \dim \Ng$ are finite and $\Mg$ and $\Ng$ are dual spaces of each 
other with respect to the inner product of $\Hg$. 
Let $S$ be the Riesz projection onto $\Mg$. 

\bglm \lblm(threshold)
\ben 
\item[{\rm (1)}] $aD_0 a$ is an isomorphism from $\Mg$ onto $\Ng$ 
and $bD_0 b$ from $\Ng$ onto $\Mg$. They are inverses of each other. 
\item[{\rm (2)}] $(a\ph, D_0 a\ph)$ is an inner product on $\Mg$ 
and $(b\p, D_0 b\p)$ on $\Ng$.
\item[{\rm(3)}] For an orthonormal basis $\{\ph_1, \dots, \ph_n\}$ 
of $\Mg$ with respect to the inner product 
$(a\ph, D_0 a\ph)$, define $\p_j= a D_0 a \ph_j$, $j=1, \dots, n$. 
Then: 
\ben 
\item[{\rm (a)}] $\{\p_1, \dots, \p_n\}$ is an orthonormal basis of $\Ng$ 
with respect to $(b\p, D_0 b\p)$.
\item[{\rm (b)}] $\{\ph_1, \dots, \ph_n\}$ and 
$\{\p_1, \dots, \p_n\}$ are dual basis of $\Mg$ and $\Ng$ 
respectively. 
\item[{\rm (c)}] 
$S f = \la f, \p_1 \ra \ph_1+ \cdots+ \la f, \p_n \ra \ph_n$, $f \in \Hg$. 
\een
\een
\edlm 
\bgpf (1) Let $\ph \in \Mg$. Then, $\ph=-bD_0 a\ph$ and  
$aD_0a \ph = - aD_0 b \cdot a D_0 a\ph$. Hence 
$aD_0 a\ph\in \Ng$. Likewise $bD_0 b$ maps $\Ng$ into $\Mg$.
We have  
\begin{gather*}
bD_0 b\cdot aD_0a \ph= (bD_0a)^2 \ph =\ph, \quad \ph\in \Mg, \\
aD_0a \cdot bD_0b \p= (aD_0 b)^2 \p=\p, \quad \p \in \Ng 
\end{gather*}
and $aD_0 a$ and $bD_0 b$ are inverses of each other. \\[5pt]
(2) Let $\ph \in \Mg$. Then $a\ph \in L^1 \cap L^{\s}$ for 
some $\s>3/2$ (see the proof of \reflm(eigen-or-reso) below) 
and $\widehat{a\ph} \in L^\infty \cap L^\r$ 
for some $\r<3$ by Hausdorff-Young's inequality. It follows that   
\[
(a\ph, D_0 a\ph)= \int_{\R^3} \frac{|\widehat{a\ph}(\xi)|^2}{|\xi|^2}d\xi 
\geq 0  
\]
and $(a\ph, D_0 a\ph)=0$ implies $a\ph=0$ hence, $\ph=-bD_0 a \ph=0$. 
Thus, $(a\ph, D_0 a\ph)$ is an inner product  of $\Mg$. 
The proof for $(b\p, D_0 b\p)$ is similar. 
\\[5pt]
(3) We have for any $j,k=1, \dots, n$ that 
\[
(b\p_j, D_0 b\p_k)=(b aD_0 a \ph_j, D_0 b aD_0 a \ph_k)
=(-a\ph_j, -D_0 a \ph_k)=\d_{jk} 
\]
and $\{\p_1, \dots, \p_n\}$ is orthonormal with respect to the inner 
product $(b\p, D_0 b\p)$. Since $n= \dim \Ng$, it is a basis of $\Ng$. 
\[
(\ph_j, \p_k)= 
(\ph_j, a D_0 a \ph_k)=(a\ph_j, D_0 a \ph_k)=\d_{jk}, \quad j,k=1, \dots, n.
\]
Hence $\{\ph_j\}$ and $\{\p_k\}$ are dual basis of each other. 
Because of this, (c) is a well known fact for Riesz projections 
to eigen-spaces of compact operators (\cite{Kato}). 
This completes the proof of the lemma.
\edpf 

The following lemma should be known for a long time.  
We give a proof for readers' convenience.

\bglm \lblm(HD-type)
Let $1<\c\leq 2$ and $\s<3/2<\r$. Then, the integral operator  
\bqn \lbeq(15)
(\Qg_\c u)(x) = \int_{\R^3}\frac{\ay^{-\c}u(y)}{|x-y|}dy 
\eqn 
is bounded from $(L^\s\cap L^\r)(\R^3)$ to the space $C_\ast(\R^3)$ 
of bounded continuous functions on $\R^3$ which converge to $0$ as 
$|x|\to 0$:  
\bqn \lbeq(16)
\|\Qg_\c u\|_{L^\infty} \leq C \|u\|_{(L^\s \cap L^\r)(\R^3)}.
\eqn 
For $R \geq 1$, there exists a constant 
$C$ independent of $u$ such that for $|x|\geq R$
\bqn \lbeq(lm)
\left| (Q_{\c}u)(x) - \frac{C(u)}{|x|}\right| \leq  
C\frac{\|u\|_{L^\s\cap L^\r}}{\ax^{\c}}, \quad 
C(u)=\int_{\R^3} \ay^{-\c}u(y)dy.
\eqn 
\edlm 
\bgpf We omit the index $\c$ in the proof. 
Since $|x|^{-1}\in L^{3,\infty}(\R^3)$, it is obvious that 
$\Qg u(x)$ is a bounded continuous function 
and that \refeq(16) is satisfied. Thus, it suffices to prove 
\refeq(lm) for $|x|\geq 100$. 
Let $K_{x}$ be the unit cube with center $x$.
Combining the two integrals on the left hand side of \refeq(lm), 
we write it as 
\[
(Q_{\c}u)(x) - \frac{C(u)}{|x|}= \frac1{|x|}\left(
\int_{K_{x}} + \int_{\R^3 \setminus K_{x}}\right)
\frac{(2yx-y^2)\ay^{-\c}u(y)}{|x-y|(|x-y|+|x|)}dy 
\equiv I_0(x) + I_1(x).
\]
When $|x-y|\leq 1$ and $|x|\geq 100$, $|x|, \ax, |y|$ and  $|x-y|$ 
are comparable in the sense that $0<C_1\leq |x|/\ax \leq C_2<\infty$ 
and etc. and we may estimate the integral over $K_{x}$ as follows: 
\bqn 
|I_0(x)| \leq \frac{C}{|x|\ax^{\c-1}}
\int_{K_{x}} \frac{|u(y)|}{|x-y|}dy 
\leq \frac{C}{\ax^{\c}}\|u\|_{L^\r(K_{x})}. 
\eqn 
We estimate the integral $I_1(x)$ by splitting it as  
$I_1(x)=I_{10}(x) + I_{11}(x)$:
\begin{gather*}
I_{10}(x)= \frac{-1}{|x|}\int_{\R^3 \setminus K_{x}}
\frac{y^2\ay^{-\c}u(y)}{|x-y|(|x-y|+|x|)}dy, \\
I_{11}(x)= \frac1{|x|}\int_{\R^3 \setminus K_{x}}
\frac{2yx\ay^{-\c}u(y)}{|x-y|(|x-y|+|x|)}dy.
\end{gather*}
Since $|x-y|+ |x|\geq C\ax^{\c-1} \ay^{2-\c}$ for $|x|\geq 100$, 
H\"older's inequality implies  
\bqn \lbeq(10)
|I_{10}(x)|\leq \frac{C}{|x| \ax^{\c-1}}\int_{\R^3\setminus K_{x}} 
\frac{|u(y)|}{|x-y|}dy 
\leq \frac{C}{\ax^{\c}} \|u\|_{L^\rho(\R^3)}.
\eqn 
Let $\s'$ be the dual exponent of $\s$. Then, $\s'>3$ and 
via H\"older's inequality  
\bqn \lbeq(11)
|I_{11}(x)| \leq C\left(
\int_{\R^3}\left(\frac{\ay^{1-\c}}
{\la x-y\ra(\ax + \ay)}\right)^{\s'}dy \right)^{1/\s'}
\|u\|_{L^\s(\R^3)}.
\eqn 
If $|x|<100|y|$ then $\ay^{\c-1}(\ax+ \ay)\geq C \ax^{\c}$ and 
\bqn \lbeq(03)
\left(
\int_{|x|<100|y|}\left(\frac{\ay^{1-\c}}
{\la x-y\ra(\ax + \ay)}\right)^{\s'}dy \right)^{1/\s'}
\leq \frac{C}{\ax^{\c}} \|\ax^{-1}\|_{L^{\s'}}
\eqn 
When $|x|>100|y|$, we may estimate for $1< \c\leq 2$ as 
\[
\frac{\ay^{1-\c}}{\la x- y \ra(|x|+|y|)}
\leq \frac{C}{\la x-y\ra \ax^{\c}}.
\]
It follows that 
\bqn \lbeq(04)
\left(
\int_{|x|>100|y|}\left(\frac{\ay^{1-\c}}
{\la x-y\ra(\ax + \ay)}\right)^{\s'}dy \right)^{1/\s'}
\leq \frac{C}{\ax^{\c}} \|\ax^{-1}\|_{L^{\s'}}.
\eqn 
Estimates \refeq(03) and \refeq(04) imply  
\bqn \lbeq(this)
|I_{11}(x)|\leq \frac{C}{\ax^{\c}}\|u\|_{L^\s}
\eqn 
Combining \refeq(this) with \refeq(10), we obtain \refeq(lm). 
\edpf  

\bglm 
\lblm(eigen-or-reso) 
\begin{enumerate} 
\item[{\rm (1)}]
The following is a continuous functional on $\Ng$: 
\[
\Ng \ni \ph \mapsto L(\ph)=\frac1{4\pi}\int_{\R^3} a(x)\ph(x)dx
=\frac1{4\pi}\la a,  \ph\ra \in \C. 
\]
\item[{\rm (2)}] For $\ph \in \Ng$, let $u=D_0 (a \ph)$. Then, 
\ben 
\item[{\rm (a)}] $u$ is a sum $u= u_1+ u_2$ of 
$u_1\in C^\infty(\R^3) \cap L^\infty(\R^3)$ and 
$u_2 \in (W^{\frac32+\ep,2}\cap W^{2,\frac32+\ep})(\R^3)$ for some 
$\ep>0$. It satisfies  
\bqn \lbeq(Seq)
(-\lap + V )u(x)=0.
\eqn
\item[{\rm (b)}] $u$ is bounded continuous and satisfies  
\bqn \lbeq(expansion)
u(x) = \frac{L(\ph)}{|x|} + O
\left(\frac{1}{|x|^{2}}
\right), \quad |x|\to \infty.
\eqn 
\item[{\rm (c)}] 
$u$ is an eigenfunction of $H$ with eigenvalue $0$ if and only if $L(\ph)=0$ 
and it is a threshold resonance of $H$ otherwise. 
\een
\item[{(3)}] The space of zero eigenfunctions in $\Ng$ has codimension 
at most one. 
\een 
\end{lemma}
\bgpf (1) Since $a \in L^2$,  
$|L(\ph)|\leq (4\pi)^{-1}\|a\|_{L^2} \|\ph\|_{L^2}$. 

\noindent 
(2a) Assumption \refeq(assump-V) implies 
$a(x) = \ax^{-1}\ta(x)$ with $\ta\in (L^{2p}\cap L^{2q})(\R^3)$ 
and $1\leq 2p<3$ and $2q>6$. It follows by H\"older's inequality that  
$\ta \ph \in L^{\frac{6}{5}-\ep} \cap L^{\frac{3}{2}+\ep}$ for 
an $\ep>0$. Using the the Fourier multiplier $\chi(D)$ 
by $\chi\in C_0^\infty(\R^3)$ such that $\chi(\xi)=1$ for $|\xi|\leq 1$,  
\[
\chi(D)u=\frac1{(2\pi)^{\frac32}}\int_{\R^3} e^{ix\xi} 
\chi(\xi)\hat u(\xi)d\xi, 
\]
we decompose $u$:   
\[
u= u_1+ u_2, \quad u_1= \chi(D)D_0 (a\ph), \ \ 
u_2=\{(1-\chi(D))(1-\lap)D_0\} (1-\lap)^{-1}(a\ph).
\]
Since $a\ph \in L^1(\R^3)$ it is obvious that 
\[
u_1(x)= \frac1{(2\pi)^{3/2}}\int_{\R^3} e^{ix\xi}\chi(\xi)\frac{u(\xi)}{|\xi|^2}
d\xi  \in C^\infty(\R^3), \quad \lim_{|x|\to \infty} \pa^\a u_1(x)=0 
\]
for all $\a$. 
Since $(1-\chi(\xi))(1+|\xi|^2)|\xi|^{-2}$ is a symbol of H\"ormander 
class $S_0$, the multiplier $(1-\chi(D))(1-\lap)D_0$ 
is bounded in any Sobolev space $W^{k,p}(\R^3)$ for $1<p<\infty$ 
by Mikhlin's theorem  and,  
\[
(1-\lap)^{-1}(a\ph) \in W^{2, \frac32+\ep}(\R^3) \cap W^{\frac32+\ep,2}(\R^3)
\]
for an $\ep>0$ by the Sobolev embedding theorem. It follows that   
$u_2 \in W^{2, \frac32+\ep}(\R^3) \cap W^{\frac32+\ep,2}(\R^3)$, 
in particular, $u$ is bounded and H\"older continuous. 
If $(1+ b D_0 a)\ph=0$, then 
\[
a(1+ bD_0 a)\ph= (1+ V D_0)a \ph =(-\lap + V)D_0 a\ph=0.
\] 
and $(-\lap+ V)u(x)=0$. 

\noindent
(2b) We just proved that $u$ is bounded and H\"older continuous.   
We use the notation in the proof of \reflm(HD-type). 
We have $a\ph = -V D_0 (a\ph) $ and 
\[
D_0 (a\ph)(x)= \frac{1}{4\pi}\left(\int_{K_{x} }+ 
\int_{\R^3 \setminus K_{x}} \right)
\frac{\ay^{-1}\ta(y)\ph(y) dy}{|x-y|} = I_1(x) + I_2(x). 
\]
Since $\ay $ is comparable with $\ax$ when $|x-y|<1$, 
\[
|I_1 (x)| \leq C \ax^{-1} \|\ta \ph\|_{L^{\frac32+\ep}} 
\||x|^{-1}\|_{L^{\t}(K_{x})}, \quad  \t= \tfrac{3+2\ep}{1+2\ep}<3
\]
For estimating the integral over $\R^3 \setminus K_{x}$, we use that 
$\ta \ph\in L^{\frac{6}{5}-\ep}$ for some $0<\ep<1/5$. 
Let $\d=(6-5\ep)/(1-5\ep)$. Then, $\d>6$ and H\"older's inequality implies 
\[
|I_2(x)| \leq C \|\ta\ph \|_{L^{\frac{6}{5}-\ep}}
\left(\int_{\R^3}\frac{dy}{\la x- y \ra^{\d}\ay^{\d}}\right)^{\frac{1}{\d}}
\leq \frac{C \|\ta \ph\|_{L^{\frac{6}{5}-\ep}}} {\ax}.  
\] 
Hence, $a\ph = - VD_0(a\ph) \in \ax^{-3} (L^{p} \cap L^{q})(\R^3)$ 
and \reflm(HD-type) with $\gamma=2$ implies statement (2b). 

\noindent 
Statements (2a) and (2b) obviously implies (2c). (3) follows from 
(1) and (2c). 
\edpf 

We distinguish following three cases: \\[5pt]
Case (a): $\Ng \cap \Ker(L)=\{0\}$. 
Then, \reflm(eigen-or-reso) implies 
$\dim \Ng=1$, $H$ has no zero eigenvalue and has 
only threshold resonances $\{u= D_0(a\ph) \colon \ph \in \Ng\}$. \\[5pt]
Case (b): $\Ng=\Ker(L)$. Then, 
$\{u= D_0 (a\ph) \colon \ph \in \Ng\}$ consists only of eigenfunctions 
of $H$ with eigenvalue $0$. \\[5pt]
Case (c): $\{0\}\subsetneqq \Ng \cap \Ker (L) \subsetneqq \Ng$. In this case 
$H$ has both zero eigenvalue and threshold resonances. 

In case (c), we take an orthonormal basis $\{\ph_1, \ph_2, \dots, \ph_n\}$ 
of $\Ng$ such that $\ph_2, \dots, \ph_n \in \Ker (L)$ and 
$\ph_1 \in \Ker(L)^\perp$ such that $L(\ph_1)>0$ which uniquely 
determines $\ph_1$.  

We study $\ep(1+M_\ep(\ep{k}))^{-1}$, 
$M_\ep(\ep{k})=\l_0(\ep)b G_0(\ep{k})a$ as $\ep \to 0$ by applying 
the following \reflm(JN) due to Jensen and Nenciu (\cite{JN}). 
We consider the case (c) only. 
The modification for the case (a) and (b) should be obvious.

\begin{lemma} \lblm(JN) 
Let $\Ag$ be a closed operator in a Hilbert space $\Hg$ 
and $S$ a projection. Suppose $\Ag+S$ has a bounded inverse. 
Then, $\Ag$ has a bounded inverse if and only if 
\[
\Bg= S - S(\Ag+S)^{-1}S 
\]
has a bounded inverse in $S\Hg$ and, in this case, 
\bqn \lbeq(JN-1)
\Ag^{-1}= (\Ag+S)^{-1}+ (\Ag+S)^{-1}S\Bg^{-1}S (\Ag+S)^{-1}.
\eqn 
\end{lemma}

We recall \refeq(luG) and \refeq(defQQ). We apply \reflm(JN) to 
\bqn \lbeq(Agdef)
\Ag=1 +M_\ep(\ep{k})\equiv  1+ \l(\ep) b G_0(\ep{k}) a . 
\eqn 
We take as $S$ the Riesz projection onto the kernel $\Mg$ of 
$Q_0= 1+ bD_0 a$. Since $bD_0 a$ is compact, 
$Q_0 + S$ is invertible. Hence, by virtue of \refeq(luG), 
$\Ag +S$ is also invertible for small $\ep>0$ and 
the Neumann expansion formula yields,  
\begin{align}
(\Ag+S)^{-1}& = (Q_0 + \ep Q_1 + O(\ep^2) + S)^{-1} \notag \\
& = 
\Big(1+ \ep (Q_0 +S)^{-1} Q_1 + O(\ep^2)\Big)^{-1}(Q_0 +S)^{-1}
\notag  \\
& = (Q_0 +S)^{-1} - 
\ep (Q_0 +S)^{-1} Q_1 (Q_0 +S)^{-1} + O(\ep^2).  \lbeq(AgS)
\end{align}
Since $S(Q_0 +S)^{-1}= (Q_0 +S)^{-1}S = S$, 
the operator $\Bg$ of \reflm(JN) corresponding  
to $\Ag$ of \refeq(Agdef) becomes 
\bqn 
\Bg = \ep S Q_1 S + O(\ep^2), \quad \sup_{k\in \W}
\|O(\ep^2)\|_{\Bb(\Hg)}\leq C \ep^2, \lbeq(3-22)
\eqn 
where $\W\Subset\Cb^{+}\setminus \{0\}$. 
Take the dual basis $(\{\ph_j\}, \{\psi_j\})$ of $(\Mg, \Ng)$ 
defined in \reflm(threshold). Then, 
$bD_0 a \ph= -\ph$ for $\ph \in \Mg$,  
$(a,\ph_j)=0$ for $2\leq j \leq n$ and 
$(\p_j, b)= (aD_0 a\ph_j, b)= -(\ph_j,a)$ imply  
\[
S Q_1 S = S(\l'(0) b D_0 a + ik b D_1 a)S 
= -\l'(0)S - \frac{ik}{4\pi} |(a,\ph_1)|^2 (\ph_1 \otimes \p_1).
\]
It follows from \refeq(3-22) that uniformly with respect to $k \in \W$ 
we have    
\bqn \lbeq(Bg)
\left\|\ep \Bg^{-1}+ \left(
\l'(0)+ i\frac{k|(a,\ph_1)|^2}{4\pi}\right)^{-1} \ph_1 \otimes \p_1 
 + \l'(0)^{-1} \sum_{j=2}^n \ph_j \otimes \p_j \right\| \leq C \ep 
\eqn 
Then,  since 
$\|(\Ag+S)^{-1}\|_{\Bb(\Hg)}$ is bounded as $\ep \to 0$ and $k \in \W$ 
and 
\[
\lim_{\ep \to 0}\sup_{k\in \W} (\|S(\Ag+S)^{-1}-S\|_{\Bb(\Hg)}+ 
\|(\Ag+S)^{-1}S-S\|_{\Bb(\Hg)}=0,  
\]
\refeq(JN-1), \refeq(AgS) and \refeq(Bg) imply the first statement 
of the following proposition. 

\bgprp \lbprp(thresh-1)
Let $N=1$ and the assumption \refeq(assump-V) be satisfied. 
Suppose that $H$ is of exceptional type at $0$ of the case (c). 
Then, with the notation of \reflm(threshold), 
uniformly with respect to $k \in \W$ in the operator norm of $\Hg$ 
we have that 
\begin{multline} 
\lim_{\ep \to 0} \ep(1+D_\ep(\ep{k}))^{-1} \\
=  -\left(
\l'(0)+ i\frac{k|(a,\ph_1)|^2}{4\pi}\right)^{-1} \ph_1 \otimes \p_1 
 -\l'(0)^{-1} \sum_{j=2}^n \ph_j \otimes \p_j\equiv \Lg  \lbeq(limit-Ag)
\end{multline}
and that 
\bqn 
\big\la a |\,\refeq(limit-Ag)\, \big| b \big\ra 
= - \left( \a- \frac{ik}{4\pi}\right)^{-1}, \quad 
\a=- \frac{\l'(0)}{|(a,\ph_1)|^2}. 
\lbeq(j-th-diago)
\eqn 
The same result holds for other cases with the following changes:
For the case {\rm (a)} replace $\ph_1$ and $\psi_1$ by $\ph$ and $\psi$ 
respectively which are normalized as $\ph_1$ and $\psi_1$ and, for 
the case {\rm (b)} set $\ph_1=\psi_1=0$.   
\edprp

\subsection{Proof of \refthb(theo-1)}

Let $\Lg_j$, $j=1, \dots, N$ be the $\Lg$ of \refeq(limit-Ag) 
corresponding to 
$H_j(\ep)=-\lap + \l_j(\ep)V_j$. Then, applying \refprp(thresh-1) to 
$H_j(\ep)$, we have 
\bqn \lbeq(Lg)
\lim_{\ep \to 0} \ep(1+D_\ep(\ep{k}))^{-1} \\
=  \oplus_{j=1}^N  \Lg_j \equiv \tLg
\eqn 
It follows by combining \reflm(E-estimate) and \refeq(Lg) that 
\bqn \lbeq(BGgA)
\lim_{\ep \to 0}\big( 1+ \ep(1+D_\ep(\ep{k})) \big)^{-1}E_\ep(\ep{k}) 
=  1+ \tLg | B \ra \hat\Gg(k) \la A| 
\eqn 

We apply the following lemma due to Deift (\cite{Deift}) to 
the right of \refeq(BGgA). 

\bglm \lblm(Deift-lemma)
Suppose that $1+ \la A |\tLg |B\ra \hat{\Gg}(k) $
is invertible in $\Bb(\C^N)$.  Then, 
$1+  \tLg |B\ra \hat{\Gg}(k)\la A |$ is also invertible 
in $\Bb(\Hg^{(N)})$ and 
\bqn \lbeq(DE)
\la A |( 1+ \tLg |B\ra \hat{\Gg}(k)\la A |)^{-1} = 
(1+ \la A | \tLg  |B \ra \hat{\Gg}(k))^{-1} \la A  | 
\eqn 
\end{lemma}
\bgpf Since $a_1, \dots, a_N \in L^2(\R^3)$, 
$|A \ra \colon \C^N \to \Hg^{(N)}$ and $\la A | \colon \Hg^{(N)} \to 
\C^N$ are both bounded operators. Then, the lemma is 
an immediate consequence of Theorem 2 of 
\cite{Deift}. 
\edpf 

For the next lemma we use the following simple lemma for matrices. Let 
\[
\Ag= \begin{pmatrix} 
W & X \\
Y & Z  
\end{pmatrix} , 
\quad 
\Bg= \begin{pmatrix} 
0 & 0 \\
0 & V 
\end{pmatrix}
\]
be matrices decomposed into blocks.

\begin{lemma} \lblm(matrix)
Suppose $V$ and $1+ V Z$ are invertible. Then, 
\[
\left( 1 + 
\begin{pmatrix} 
0 & 0 \\
0 & V 
\end{pmatrix}
\begin{pmatrix} 
W & X \\
Y & Z  
\end{pmatrix} \right)^{-1}
\]
exists and 
\bqn\lbeq(matr-iden)
\left( 1 + 
\begin{pmatrix} 
0 & 0 \\
0 & V 
\end{pmatrix}
\begin{pmatrix} 
W & X \\
Y & Z  
\end{pmatrix} \right)^{-1}
\begin{pmatrix} 
0 & 0 \\
0 & V 
\end{pmatrix}
= 
\begin{pmatrix} 
0  & 0 \\
0 & (V^{-1}+ Z)^{-1} 
\end{pmatrix}
\eqn
\edlm 
\bgpf It is elementary to see 
\begin{multline}
\left(1+ \begin{pmatrix} 
0 & 0 \\
0 & V
\end{pmatrix}
\begin{pmatrix} 
W & X \\
Y & Z  
\end{pmatrix} \right)^{-1} \\
= 
\begin{pmatrix} 
1 & 0 \\
VY & 1+ VZ  
\end{pmatrix}^{-1} = 
\begin{pmatrix} 
1 & 0 \\
-(1+ VZ)^{-1}VY &  (1+ VZ)^{-1} 
\end{pmatrix}
\end{multline}
and the left side of \refeq(matr-iden) is equal to 
\[
\begin{pmatrix} 
0 & 0 \\
0 & (1+ VZ)^{-1}V
\end{pmatrix}
= \begin{pmatrix} 
0 & 0 \\
0 & (V^{-1}+ Z)^{-1}
\end{pmatrix}
\]
which proves the lemma. 
\edpf

\bglm \lblm(f-lemma) Let $k\in \W$. Then, 
$1+ \la A|\tLg |B\ra \hat{\Gg}(k)$ is invertibe in $\C^N$. 
If $H_1, \dots, H_{N}$ are arranged in such a way that 
$H_1, \dots, H_{n_1}$ have no resonances and  
$H_{n_1+1}, \dots, H_{N}$ do and, $N=n_1+ n_2$, then 
\bqn \lbeq(final)
(1+ \la A|\tLg |B\ra \hat{\Gg}(k))^{-1}\la A|\tLg|B\ra 
= \begin{pmatrix}
{\mathbb O}_{n_1 n_1}  & {\mathbb O}_{n_1 n_2} \\
{\mathbb O}_{n_2 n_1}  & - \tilde{\Ga}(k)^{-1} 
\end{pmatrix}, 
\eqn
where ${\mathbb O}_{n_1 n_1}$ is the zero matrix of size $n_1 \times n_1$ 
and etc. and 
\bqn \lbeq(ga-def-a)
\tilde {\Ga}(k)= \Big(\Big(\a_j-\frac{ik}{\,4\pi\,}\Big)
\d_{j,\ell} - \Gg_k(y_j-y_\ell)\hat{\d}_{j\ell}
\Big)_{\!j,\ell=n_1+1,\dots,N}.
\eqn 
\edlm 
\bgpf We let $\ph_{j1}$ be the resonance of 
$H_j$, $j=n_1+1, \dots, N$, corresponding to $\ph_1$ of the previous 
section and define  
\bqn \lbeq(def-of-alpha)
\a_j = - \frac{\l'(0)}{|(a_j,\ph_{j1})|^2}. 
\eqn 
Then, \refprp(thresh-1) implies that, 
\[
\la A |\tLg| B\ra =\begin{pmatrix} 
0   &         &       &         &     & \\
    & \ddots  &        &        &     &  \\
    &         &  0     &        &     &   \\
    &         &        &  -\left(\a_{n_2+1} -\frac{ik}{4\pi}\right)^{-1} & 
    &  \\  
    &     &     &   &  \ddots &   \\ 
    &     &     &   &        & - \left(\a_{n_1+ n_2} -\frac{ik}{4\pi}\right)^{-1}
\end{pmatrix}.
\]
and we obtain \refeq(final) by applying \reflm(matrix) to the left of \refeq(final) with 
\[
V = \begin{pmatrix} 
    -\left(\a_{n_2+1} -\frac{ik}{4\pi}\right)^{-1} &     &  \\  
    &  \ddots &   \\ 
    &          & -\left(\a_{n_1+ n_2} -\frac{ik}{4\pi}\right)^{-1}
\end{pmatrix}.
\]
and with 
\[
\begin{pmatrix} 
W & X \\
Y & Z  
\end{pmatrix}  = \hat {\Gg}(k).
\]
\edpf 

\reflm(Deift-lemma) and \reflm(f-lemma) imply that the following limit 
exists in $\Bb(\Hg)$ and 
\[
\lim_{\ep \to 0} 
\big( 1+ \ep(1+D_\ep(\ep{k}))^{-1}E_\ep(\ep{k}) \big)^{-1}
=  \big(1+ \tLg | B \ra \hat \Gg(k) \la A|\big)^{-1} 
\]
and hence so does 
\bqn \lbeq(ff)
\lim_{\ep\to 0} \ep\big(1+ M_\ep(\ep{k})\big)^{-1}= 
\big(1+ \tLg | B \ra \hat \Gg(k) \la A|\big)^{-1}\tLg 
\eqn

\paragraph{Completion of the proof of \refthb(theo-1)}
By the assumption of the theorem, we may assume $n_1=0$ in 
\reflm(f-lemma). Abusing notation, we write  
\[
\hat{\Gg}_{k}^{(N)} u = (\hat{\Gg}_{k} u)^{(N)}, \quad 
\hat{\Gg}_{k}u= \frac1{4\pi}\int_{\R^3}\frac{e^{ik|x|}u(x)}{|x|}dy.
\]
We first prove \refeq(wave-converge) for the $+$ case. 
We let $u, v \in \Dg_\ast$ and $R>0$. Then, 
\refeq(7-1) and \refeq(ff) imply that 
\bqn \lbeq(LHS-a)
\ep^2 ((1+M_\ep(-\ep k))^{-1}\L(\ep) 
B (G_0 (k\ep)- G_0(-k\ep))^{(N)}U_\ep u, A G_0(k\ep)^{(N)}  U_\ep v ) 
\eqn 
converges as $\ep \to 0$ to 
\bqn \lbeq(Ng)
( \la A | (1+ \tLg |B \ra \hat{\Gg}(-k) 
\la A| )^{-1} \tLg |B\ra \la (\Gg_{k}^{(N)}-\Gg_{-k}^{(N)})u, 
\Gg_{k}^{(N)}v ) 
\eqn 
uniformly with respect to $k\in [R^{-1}, R]$. Here we have 
\begin{multline}
\la A | (1+ \tLg |B \ra \hat{\Gg}(-k) \la A| )^{-1} \tLg |B\ra 
\\ 
= (1+ \la A| \Lg |B \ra \hat{\Gg}(-k))^{-1} 
\la A|\Lg|B \ra = - \tilde{\Ga}(-k)^{-1} \lbeq(once)
\end{multline}
by virtue of \refeq(DE) and \refeq(final). Thus, \refeq(LHS-a) 
converges as $\ep \to 0$ to 
\[
- (\Ga_{\alpha,Y}(-k)^{-1}
\big(\hat{\Gg}_{k}- \hat{\Gg}_{-k}\big)^{(N)}u, \hat{\Gg}_{k}^{(N)}v)
\]
uniformly on $[R^{-1}, R]$. Thus, replacing $u$ and $v$ respectively 
by ${\tau} u$ and ${\tau} {v}$, we obtain 
$W_{Y,\ep}^{+}\to W^{+}_{\a, Y}$  
strongly as $\ep \to 0$ in view of \refeq(Omega_jk) and \refeq(fr-w). 

By virtue of  \refeq(resolvent_identity) and \refeq(fr-w1),   
for proving the convergence \refeq(epto0) of the resolvent, 
it suffices to show that as $\ep \to 0$ in the strong topology of $\Bb(\Hg)$ 
\begin{multline} \lbeq(MO)
\ep^{2}U_\ep G_0(k\ep)^{(N)}A(1+M_\ep(\ep k))^{-1}\L(\ep) 
{\ep}B G_0 (k\ep)^{(N)}U_\ep \\
\to - |\hat{\Gg}_k^{(N)} \ra \Ga_{\alpha,Y}(k)^{-1} \la \hat{\Gg}_k^{(N)}| 
\end{multline} 
for every $k \in \C^{+}\setminus \Eg$. 
However, \refeq(7-1), \refeq(7-1a) and \refeq(ff) imply that for 
$k \in \C^{+}\setminus \Eg$ the first line of \refeq(MO) 
converges strongly in $\Bb(\Hg)$ as $\ep \to 0$ to  
\bqn 
|\Gg_k^{(N)}\ra \la A | (1+ \tLg |B \ra \hat{\Gg}(k) 
\la A| )^{-1} \tLg |B\ra \la \Gg_{k}^{(N)}|.  \lbeq(Ng-a)
\eqn 
This is equal to the second line by virtue of \refeq(once)  
with $k$ in place of $-k$. This completes the proof of the theorem.

\parbox{8.5cm}{\it \small Artbazar Galtbayar \\
Center of Mathematics for Applications and \\
Department of Applied Mathematics\\
National University of Mongolia \\
University Street 3,\\
 Ulaanbaatar, (Mongolia)\\}

\parbox{8.5cm}{\it \small Kenji Yajima \\
Department of Mathematics \\
Gakushuin University\\
1-5-1 Mejiro\\
Toshima-ku
Tokyo 171-8588 (Japan).} 
\end{document}